\documentclass[twocolumn,superscriptaddress,letter]{revtex4}%
\usepackage{amssymb}
\usepackage{color}
\usepackage{graphicx}
\usepackage{dcolumn}
\usepackage{bm}
\usepackage[header,title,page,titletoc]{appendix}
\usepackage{amsmath}
\usepackage{amsfonts}%
\providecommand{\U}[1]{\protect \rule{.1in}{.1in}}

\begin{document}
\title{Irrational Non-Abelian Statistics for Non-Hermitian Generalization of Majorana
Zero Modes}
\author{Xiao-Ming Zhao}
\affiliation{Beijing National Laboratory for Condensed Matter Physics, Institute of
Physics, Chinese Academy of Sciences, Beijing 100190, China}
\affiliation{Center for Advanced Quantum Studies, Department of Physics, Beijing Normal
University, Beijing 100875, China}
\author{Cui-Xian Guo}
\affiliation{Beijing National Laboratory for Condensed Matter Physics, Institute of
Physics, Chinese Academy of Sciences, Beijing 100190, China}
\affiliation{Center for Advanced Quantum Studies, Department of Physics, Beijing Normal
University, Beijing 100875, China}
\author{Meng-Lei Yang}
\affiliation{Center for Advanced Quantum Studies, Department of Physics, Beijing Normal
University, Beijing 100875, China}
\author{Heng Wang}
\affiliation{Center for Advanced Quantum Studies, Department of Physics, Beijing Normal
University, Beijing 100875, China}
\author{Wu-Ming Liu}
\affiliation{Beijing National Laboratory for Condensed Matter Physics, Institute of
Physics, Chinese Academy of Sciences, Beijing 100190, China}
\author{Su-Peng Kou}
\thanks{Corresponding author}
\email{spkou@bnu.edu.cn}
\affiliation{Center for Advanced Quantum Studies, Department of Physics, Beijing Normal
University, Beijing 100875, China}

\begin{abstract}
In condensed matter physics, non-Abelian statistics for Majorana zero modes
(or Majorana Fermions) is very important, really exotic, and completely
robust. The race for searching Majorana zero modes and verifying the
corresponding non-Abelian statistics becomes an important frontier in
condensed matter physics. In this letter, we generalize the Majorana zero
modes to non-Hermitian (NH) topological systems that show universal but quite
different properties from their Hermitian counterparts. Based on the NH
Majorana zero modes, the orthogonal and nonlocal Majorana qubits are well
defined.
In particular, due to the particle-hole-symmetry breaking, NH
Majorana zero modes have irrational non-Abelian statistics with continuously
tunable braiding Berry phase from $\pi/8$ to $3\pi/8$. This is
quite different from the usual non-Abelian statistics with fixed braiding
Berry phase $\pi/4$ and becomes an example of ``irrational topological
phenomenon". The one-dimensional NH Kitaev model is taken as an example to
numerically verify the irrational non-Abelian statistics for two NH Majorana
zero modes. The numerical results are exactly consistent with the theoretical
prediction. With the help of braiding these two zero modes, the $\pi/8$ gate
can be reached and thus universal topological quantum computation becomes possible.

\end{abstract}
\maketitle

Majorana zero modes (MZMs) have recently attracted much attention due to their
potential application in topological quantum computations (TQCs)
\cite{read2000,kitaev2001,kitaev1997,Ivanov2001,Sarma2006,Tewari2007,Nayak2008,Fu2008,Stern2010,Sau2010,Alicea2011,Mourik2012,Deng2012,Rokhinson2012,Alicea2012,Mebrahtu2013,Nadj-Perge2014,Lee2014}%
. MZMs have been predicted to be induced by vortices in a two-dimensional (2D)
spinless $p_{x}+ip_{y}$-wave superconductor (SC) \cite{read2000}, or localized
at the ends of a one-dimensional (1D) $p$-wave SC \cite{kitaev2001}. For these
topological superconductors (TSCs) with MZMs, topologically protected
degenerate ground states (referred to as Majorana qubits) exist. Based on
braiding these MZMs that obey non-Abelian statistics
\cite{Ivanov2001,Alicea2011}, a TQC is proposed \cite{Tewari2007,Nayak2008}.
Unfortunately, because the $\pi/8$ gate cannot be reached by braiding
processes, a universal TQC based on MZMs has become unrealistic and still
remains a challenge.

Alternatively, in recent years non-Hermitian (NH) physics has become an active
research areas that has attracted considerable research
\cite{Rudner2009,Esaki2011,Hu2011,Liang2013,Zhu2014,Lee2016,San2016,Leykam2017,Shen2018,Lieu2018,
Xiong2018,Kawabata2018,Gong2018,Yao2018,YaoWang2018,Yin2018,Borgnia2020,Kunst2018,KawabataUeda2018,Alvarez2018,
Jiang2018,Ghatak2019,Avila2019,Jin2019,Lee2019,Liu2019,38-1,38,chen-class2019,Edvardsson2019,
Herviou2019,Yokomizo2019,zhouBin2019,Kunst2019,Deng2019,SongWang2019,xi2019,Longhi2019,chen-edge2019,Zeuner2015,Weimann2017,Xiao2017,Bandres2018,Zhou20182,Cerjan2019,Wang2019,Xiaoxue2019,Helbig2019,Wang2015,Yuce2016,Zeng2016,Menke2017,Li2018,Lieu2019,Ashida2020}%
. Researchers have investigated some NH effects on topological SCs and MZMs.
Previous work has focus mainly on two types of NH terms on TSCs: gain/loss in
SCs induced by imaginary chemical potentials
\cite{Wang2015,Yuce2016,Zeng2016,Menke2017,Kawabata2018,Lieu2019} and
imbalanced pairing \cite{Li2018}, where the MZMs show similar properties as to
those in a Hermitian system. However, many open questions still exist
regarding the MZMs in NH TSCs:

\emph{1) Can we generalize the MZMs to NH systems that show universal but
quite different properties from their Hermitian counterparts? }

\emph{2) Can the NH effect change the non-Abelian statistics of MZMs? }

\emph{3) Do NH MZMs provide an alternative approach to universal TQC beyond
their Hermitian counterparts?}

In this letter, we aim to answer the above questions and develop a theory for
the NH generalized MZMs and the corresponding NH generalization for
non-Abelian statistics (referred to as \emph{irrational non-Abelian
statistics}).

\textbf{Non-Hermitian Majorana zero modes. }In certain TSCs, MZMs always
emerge around defects (for example, the quantized vortex in 2D TSCs or the end
in 1D TSCs). In general, a single MZM (sometimes referred to as Majorana
fermion for Hermitian TSCs) can be described by a real fermionic field
$\gamma,$ i.e., $\gamma^{\dagger}$$=$$\gamma$. We can label two MZMs by
complex fermions as $\gamma_{1}$$=$$c_{1}+c_{1}^{\dagger}$, $\gamma_{2}$%
$=$$-i(c_{2}-c_{2}^{\dagger})$ and use them to represent the basis states of a
non-local Majorana qubit:
\begin{equation}
\left \vert 0\right \rangle _{\text{M}}\equiv \frac{1}{\sqrt{2}}(\left \vert
\overline{11}\right \rangle +\left \vert \overline{00}\right \rangle ),\text{
}\left \vert 1\right \rangle _{\text{M}}\equiv \frac{1}{\sqrt{2}}(\left \vert
\overline{10}\right \rangle +\left \vert \overline{01}\right \rangle ),
\end{equation}
where $\left \vert \overline{mn}\right \rangle =\left \vert \overline
{m}\right \rangle _{1}\otimes \left \vert \overline{n}\right \rangle _{2}$ with
$m,n=0,1$, $(\left \vert \overline{0}\right \rangle _{i},\left \vert \overline
{1}\right \rangle _{i})=(\left \vert \overline{0}\right \rangle _{i}%
,c_{i}^{\dagger}\left \vert \overline{0}\right \rangle _{i})$ are the
eigenstates for complex fermions $c_{i}^{\dagger}$, $i=1,2$. In addition,
$\left \vert 0\right \rangle _{\text{M}}$ is a fermion-empty state, $\left \vert
1\right \rangle _{\text{M}}=C_{\text{M}}^{\dagger}\left \vert 0\right \rangle
_{\text{M}}$ is a fermion-occupied state where $C_{\text{M}}^{\dagger}$ is the
composite fermionic operator (see Appendix-A), $C_{\text{M}}^{\dagger}%
=(\gamma_{1}-i\gamma_{2})/2=(c_{1}+c_{1}^{\dagger}-c_{2}+c_{2}^{\dagger})/2.$
The fermion parities for the two states of the Majorana qubit are different:
the fermion parity of $\left \vert 0\right \rangle _{\text{M}}$ is even and the
fermion parity of $\left \vert 1\right \rangle _{\text{M}}$ is odd . By
introducing the fermionic parity operator $\hat{P}_{F}=(-1)^{\sum_{j}%
c_{j}^{\dagger}c_{j}},$ we have $\hat{P}_{F}\left \vert 0\right \rangle
_{\text{M}}=\left \vert 0\right \rangle _{\text{M}}$ and $\hat{P}_{F}\left \vert
1\right \rangle _{\text{M}}=-\left \vert 1\right \rangle _{\text{M}}$ .
\begin{figure}[ptb]
\includegraphics[clip,width=0.45\textwidth]{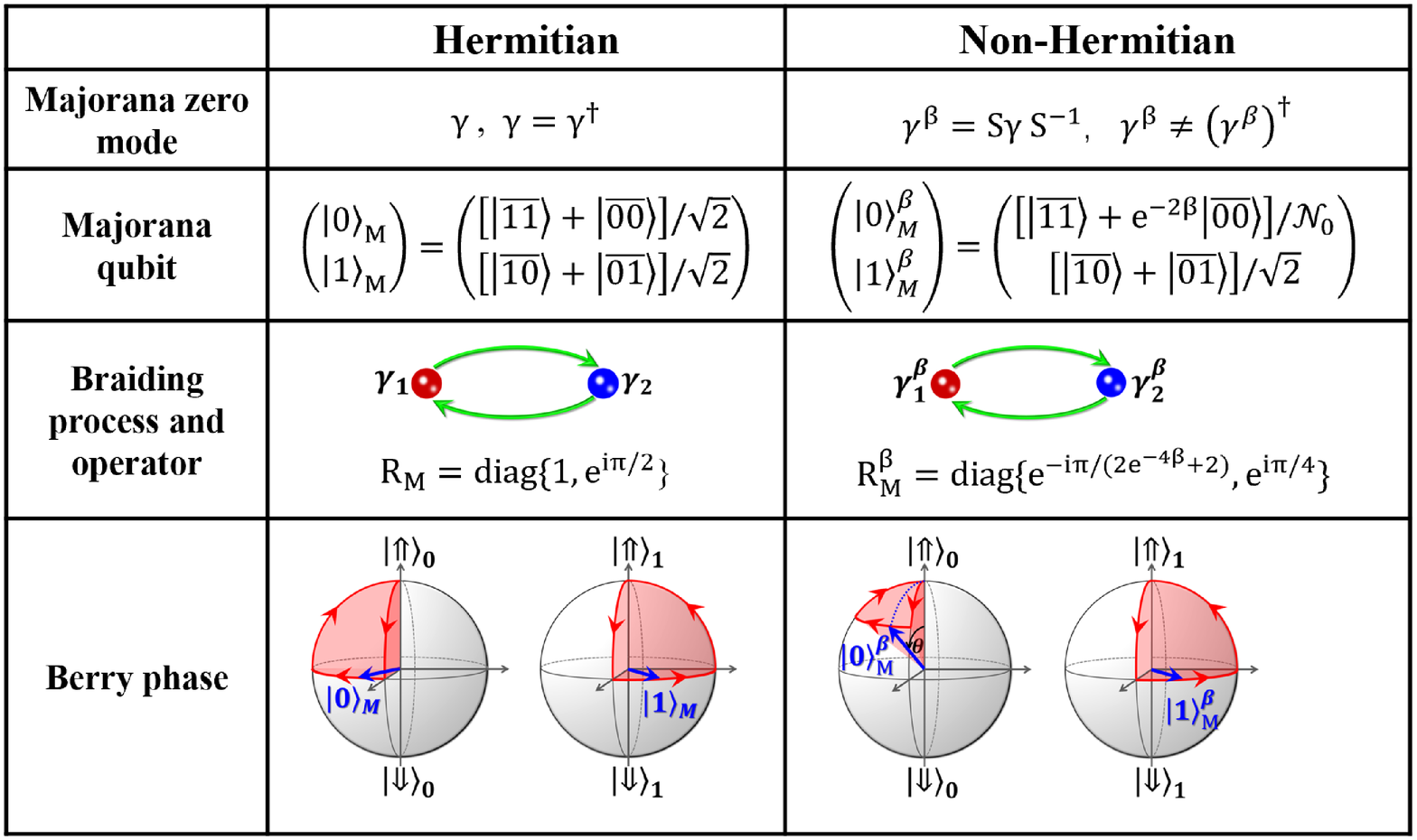}\caption{ An illustration
to show the comparison between the typical Majorana qubits and NH Majorana
qubits. We list the MZMs, orthogonality of the qubit states, braiding operator
and Berry phase in the row 2-5, respectively.}%
\end{figure}

However, \emph{can we generalize the MZMs and the corresponding Majorana qubit
to NH systems}? The answer is \emph{yes}! For a Hermitian system, a global
phase transformation $\mathcal{S}$ for fermion operators is defined as
\begin{equation}
(c,c^{\dagger})\mapsto \mathcal{S}(c,c^{\dagger})\mathcal{S}^{-1}=(e^{-i\phi
}c,e^{i\phi}c^{\dagger})
\end{equation}
with a real $\phi$. Here, we generalize the phase $\phi$ from a real number to
an imaginary number $\phi=-i\beta$, and the imaginary phase transformation
becomes a NH particle-hole (PH) similarity transformation, i.e.,
$(c,c^{\dagger})\mapsto(e^{-\beta}c,e^{\beta}c^{\dagger})$ with real $\beta$.
Therefore, with the help of the NH PH similarity transformation $\mathcal{S}$,
we define the NH MZMs as $\gamma_{i}^{\beta}=\mathcal{S}\gamma_{i}%
\mathcal{S}^{-1},i=1,2$ and we have $\gamma_{1}^{\beta}=e^{-\beta}%
c_{1}+e^{\beta}c_{1}^{\dagger},$ $\gamma_{2}^{\beta}=-i(e^{-\beta}%
c_{2}-e^{\beta}c_{2}^{\dagger}),$ where the NH strength $\beta$ is a real
number ($\beta=\beta^{\ast}\neq0$). In particular, the NH MZMs satisfy
$(\gamma_{i}^{-\beta})^{\dagger}=\gamma_{i}^{\beta}$, $(\gamma_{i}^{\beta
})^{\dagger}\neq \gamma_{i}^{\beta}$. Therefore, the properties of NH MZMs are
characterized by $\beta$. The NH PH similarity transformation $\mathcal{S}$
breaks intrinsic PH symmetry in a TSC, i.e., the symmetry between $c$ and
$c^{\dagger}$ is broken. The corresponding TSCs with NH MZMs are no longer
Hermitian and the corresponding operators $\gamma^{\beta}$ are no longer real.

We consider a TSC with two NH MZMs $\gamma_{1}^{\beta},$ $\gamma_{2}^{\beta}$
and the corresponding fermionic operators are defined as $\tilde
{C}_{\mathrm{M}}^{\dagger}=(\gamma_{1}^{\beta}-i\gamma_{2}^{\beta})/2$ and
$\tilde{C}_{\mathrm{M}}=(\gamma_{1}^{\beta}+i\gamma_{2}^{\beta})/2$, with $\{
\tilde{C}_{\mathrm{M}},\tilde{C}_{\mathrm{M}}^{\dagger}\}=1,$ $(\tilde
{C}_{\mathrm{M}})^{2}=(\tilde{C}_{\mathrm{M}}^{\dagger})^{2}=0$. We therefore
introduce a NH Majorana qubit $(\left \vert 0\right \rangle _{\text{M}}^{\beta
},\left \vert 1\right \rangle _{\text{M}}^{\beta})=(\left \vert 0\right \rangle
_{\text{M}}^{\beta},\tilde{C}_{\mathrm{M}}^{\dagger}\left \vert 0\right \rangle
_{\text{M}}^{\beta})$ based on the NH MZMs, which can be derived from a
Hermitian case under a global NH PH similarity transformation $\mathcal{S}$:%
\begin{equation}
\left \vert 0\right \rangle _{\text{M}}^{\beta}=\mathcal{S}\left \vert
0\right \rangle _{\text{M}},\text{ }\left \vert 1\right \rangle _{\text{M}%
}^{\beta}=\mathcal{S}\left \vert 1\right \rangle _{\text{M}}.
\end{equation}
From the definition of the NH MZMs, the energy difference between $\left \vert
0\right \rangle _{\text{M}}^{\beta}$\ and $\left \vert 1\right \rangle
_{\text{M}}^{\beta}$ disappears. Thus, there is almost no coupling between
$\gamma_{1}^{\beta}$ and $\gamma_{2}^{\beta}$.

For the NH Majorana qubits, according to $\mathcal{S}\hat{P}_{F}%
\mathcal{S}^{-1}=\hat{P}_{F}$, the fermion parity is also a good quantum
number and the fermion parity for the NH qubits are the same as their
Hermitian counterpart: $\hat{P}_{F}\left \vert 0\right \rangle _{\text{M}%
}^{\beta}=\left \vert 0\right \rangle _{\text{M}}^{\beta},$ $\hat{P}%
_{F}\left \vert 1\right \rangle _{\text{M}}^{\beta}=-\left \vert 1\right \rangle
_{\text{M}}^{\beta}.$ In addition, we emphasize that the PH-symmetry for the
``empty'' state $\left \vert 0\right \rangle _{\text{M}}^{\beta}$ is broken, but
the PH-symmetry for the ``occupied'' state $\left \vert 1\right \rangle
_{\text{M}}^{\beta}$ is unbroken. Under PH transformation, we have
\begin{equation}
\left \vert 0\right \rangle _{\text{M}}^{\beta}\mapsto \left \vert 0^{\prime
}\right \rangle _{\text{M}}^{\beta}\neq \left \vert 0\right \rangle _{\text{M}%
}^{\beta}, \left \vert 1\right \rangle _{\text{M}}^{\beta}\mapsto \left \vert
1^{\prime}\right \rangle _{\text{M}}^{\beta}=\left \vert 1\right \rangle
_{\text{M}}^{\beta}.
\end{equation}
The PH-symmetry breaking of the NH Majorana qubit plays an important role in
changing the typical non-Abelian statistics to anomalous non-Abelian statistics.

\textbf{Irrational non-Abelian statistics.} First, we summarize the quantum
properties of MZMs in Hermitian cases. MZMs obey \textrm{SU(2)} level-2
non-Abelian statistics. On the one hand, the fusion rule of MZMs is given by
$\sigma \times \sigma=\mathbf{1}+\psi,$ $\psi \times \psi=\mathbf{1},$ $\psi
\times \sigma=\sigma,$ where $\mathbf{1}$ is a vacuum sector, $\psi$ is the
(complex) fermion sector, and $\sigma$ is the MZM sector. Two $\sigma
$-particles (MZMs) may either annihilate to the vacuum or fuse into a $\psi
$-particle; On the other hand, if we exchange two MZMs ($\gamma_{1},$
$\gamma_{2}$), the result of the braiding is $\gamma_{1}\rightarrow-\gamma
_{2},\text{ }\gamma_{2}\rightarrow \gamma_{1}$ and the exchange operator (the
braiding operator) $\mathcal{R}_{\text{M}}$ can be described by $\mathcal{R}%
_{\text{M}}=e^{i\frac{\pi}{4}\gamma_{1}\gamma_{2}}.$ We may call
$\mathcal{R}_{\text{M}}$ to be Ivanov's braiding operator \cite{Ivanov2001}.
During the braiding process, the Berry phases for ${\left \vert 0\right \rangle
}_{\text{M}}$ and $\left \vert 1\right \rangle _{\text{M}}$ are $0$ and $\pi/2$,
respectively. So for the Majorana qubit $(\left \vert 0\right \rangle
_{\text{M}},\left \vert 1\right \rangle _{\text{M}})$, the braiding operator is
obtained as $\mathcal{R}_{\text{M}}=\mathrm{exp}\{-i\Delta \Phi \tau_{z}\}$
which is the \emph{Ivanov's braiding operator}. Here, $\tau_{z}$ denotes a
Pauli matrix on the Majorana qubit $(\left \vert 0\right \rangle _{\text{M}%
},\left \vert 1\right \rangle _{\text{M}})$. According to the topological
feature of \textrm{SU(2)} level-2 non-Abelian statistics, the Berry phase
during braiding processes $\Delta \Phi$ is fixed to be $\pi/4$ that cannot be changed.

However, for the NH MZMs $\gamma^{\beta}$, their non-Abelian statistics are
different from the Hermitian case and become a new type of non-Abelian
statistics, namely, \emph{irrational non-Abelian statistics}.

On the one hand, there exists a typical fusion rule for the NH MZMs:
$\sigma^{\beta}\times \sigma^{\beta}=\mathbf{1}^{\beta}+\psi^{\beta},\text{
}\psi^{\beta}\times \psi^{\beta}=\mathbf{1}^{\beta},\text{ }\psi^{\beta}%
\times \sigma^{\beta}=\sigma^{\beta},$ where $\mathbf{1}^{\beta}$ is the NH
vacuum sector, $\psi^{\beta}$ is the NH (complex) fermion sector, and
$\sigma^{\beta}$ is the NH MZM sector. Two NH $\sigma^{\beta}$-particles may
either annihilate to the NH vacuum $\mathbf{1}^{\beta}$ or fuse into a NH
$\psi^{\beta}$-particle.

On the other hand, anomalous braiding processes exist for the NH MZMs
$\gamma^{\beta}$. According to the case with two NH MZMs $\gamma_{1}^{\beta}$
and $\gamma_{2}^{\beta}$, two degenerate quantum states always exist.
Consequently, the braiding process for the NH MZMs is also defined by
$\gamma_{1}^{\beta}\rightarrow-\gamma_{2}^{\beta},\text{ }\gamma_{2}^{\beta
}\rightarrow \gamma_{1}^{\beta}.$ Then, a question is: \emph{can the braiding
operator for NH MZMs }$\mathcal{R}_{\text{M}}^{\beta}$ \emph{be derived by
performing a similarity transformation on the braiding operator for the
Hermitian MZMs} $\mathcal{R}_{\text{M}}?$ The answer is \emph{no}$,$ i.e.,
$\mathcal{R}_{\text{M}}^{\beta}\neq \mathcal{SR}_{\text{M}}\mathcal{S}%
^{-1}=e^{-i\frac{\pi}{4}\tau_{z}}.$

To show why, let us derive the braiding matrix $\mathcal{R}_{\text{M}}^{\beta
}$ on the Majorana qubit during the braiding processes. The Berry phases for
the quantum states of the Majorana qubits $\left \vert 0\right \rangle
_{\text{M}}^{\beta}$ and $\left \vert 1\right \rangle _{\text{M}}^{\beta}$ from
the braiding operation are calculated by the Wilson loop method,
\begin{equation}
\left \vert A_{i}\right \vert e^{\Delta \phi_{{i}}^{\beta}}={{\prod}%
_{n=0}^{n=N_{s}}}\text{ }_{\text{M}}^{\beta}\left \langle i(\theta,\varphi
_{n})|i(\theta,\varphi_{n+1})\right \rangle _{\text{M}}^{\beta}, \label{Wloop}%
\end{equation}
where i=0,1, the amplitude $\left \vert A_{i}\right \vert =1$ when the evolution
step number $N_{s}$ is sufficiently large, and $\left \vert i(\theta
,\varphi_{n})\right \rangle _{\text{M}}^{\beta}$ is the $i$-th state of the
Majorana qubit at the $n$-step during the braiding process which is labeled by
the two parameters $\theta=2\arctan(e^{-2\beta})$ and $\varphi_{n}.$ In
particular, we have%
\begin{align}
\left \vert 0(\theta,\varphi_{n})\right \rangle _{\text{M}}^{\beta}  &
=\frac{1}{\mathcal{N}_{0}}[e^{i\varphi_{n}}\left \vert \overline{11}%
\right \rangle +e^{-2\beta}\left \vert \overline{00}\right \rangle ],\nonumber \\
\left \vert 1(\theta,\varphi_{n})\right \rangle _{\text{M}}^{\beta}  &
=\frac{1}{\sqrt{2}}[\left \vert \overline{10}\right \rangle +e^{-i\varphi_{n}%
}\left \vert \overline{01}\right \rangle ].
\end{align}
where $\mathcal{N}_{0}=\sqrt{1+e^{-4\beta}}$ is the self-normalization
coefficient of the state $\left \vert 0(\theta,\varphi_{n})\right \rangle
_{\text{M}}^{\beta}$.

First, we derive the effects of the braiding operator $\mathcal{R}_{\text{M}%
}^{\beta}$ on the quantum state $\left \vert 0\right \rangle _{\text{M}}^{\beta
}$. We map the states ($\left \vert \overline{00}\right \rangle ,\left \vert
\overline{11}\right \rangle $) onto a pseudo-spin ($\left \vert \Uparrow
\right \rangle _{0}$,$\left \vert \Downarrow \right \rangle _{0}$) and use the
Bloch sphere to label the quantum states. In the Hermitian case $\beta=0$
(Fig.1), the initial state is $\left \vert 0\right \rangle _{\text{M}}^{\beta
}=\left \vert \Uparrow \right \rangle _{0}+\left \vert \Downarrow \right \rangle
_{0}$, which is denoted by a spot at the equator on the Bloch sphere
$[\theta,\varphi]=[\pi/2,0]$. During braiding process, $\left \vert
0\right \rangle _{\text{M}}^{\beta}$ adiabatically deforms into $(e^{i\varphi
_{n}}\left \vert \Uparrow \right \rangle _{0}+\left \vert \Downarrow \right \rangle
_{0})$ and finally changes into $(e^{i\pi/2}\left \vert \Uparrow \right \rangle
_{0}+\left \vert \Downarrow \right \rangle _{0})$ denoted by another spot
$[\theta,\varphi]=[\pi/2,\pi/2]$. So the geometry phase (Berry phase) is
$\Delta \varphi(1-\cos \theta)/2$ where $\theta=\pi/2$ and $\Delta \varphi=\pi
/2$; While, in the NH case $\beta \neq0$ (Fig.1), the initial state becomes
$(\left \vert \Uparrow \right \rangle _{0}+e^{-2\beta}\left \vert \Downarrow
\right \rangle _{0})$, which is denoted by a spot away from the equator of the
Bloch sphere, $[\theta,\varphi]=[2\arctan(e^{-2\beta}),0]$. During the
braiding proposes, it adiabatically deforms into $(e^{i\varphi_{n}}\left \vert
\Uparrow \right \rangle _{0}+e^{-2\beta}\left \vert \Downarrow \right \rangle _{0}%
$, and finally changes into $(e^{i\pi/2}\left \vert \Uparrow \right \rangle
_{0}+e^{-2\beta}\left \vert \Downarrow \right \rangle _{0})$ denoted by
$[\theta,\varphi]=[2\arctan(e^{-2\beta}),\pi/2]$. After the braiding processes
we obtain the geometry phase as $\Delta \varphi(1-\cos \theta)/2$ where
$\tan(\theta/2)=e^{-2\beta}$ and $\Delta \varphi=\pi/2$.

Second, using a similar operation on $\left \vert 1\right \rangle _{\text{M}%
}^{\beta}$, we map the qubit ($\left \vert \overline{01}\right \rangle
,\left \vert \overline{10}\right \rangle $) onto a pseudo-spin ($\left \vert
\Uparrow \right \rangle _{1}$,$\left \vert \Downarrow \right \rangle _{1}$),as
shown in Fig.1(b). The braiding operators for the Majorana qubit ($\left \vert
0\right \rangle _{\text{M}}^{\beta},\left \vert 1\right \rangle _{\text{M}%
}^{\beta}$) are obtained as (see the Appendix-D3)
\begin{equation}
\mathcal{R}_{\text{M}}^{\beta}\left(
\begin{array}
[c]{c}%
\left \vert 0\right \rangle _{\text{M}}^{\beta}\\
\left \vert 1\right \rangle _{\text{M}}^{\beta}%
\end{array}
\right)  =\left(
\begin{array}
[c]{cc}%
e^{i\Delta \phi_{0}^{\beta}} & 0\\
0 & e^{i\Delta \phi_{1}^{\beta}}%
\end{array}
\right)  \left(
\begin{array}
[c]{c}%
\left \vert 0\right \rangle _{\text{M}}^{\beta}\\
\left \vert 1\right \rangle _{\text{M}}^{\beta}%
\end{array}
\right)  ,
\end{equation}
where the Berry phases are obtained as $\Delta \phi_{{0}}^{\beta}=-\frac{\pi
}{2(e^{-4\beta}+1)}$ and $\Delta \phi_{1}^{\beta}=\pi/4$. It is obvious that
the Berry phase for $\left \vert 0\right \rangle _{\text{M}}^{\beta}$ is
different from ${\left \vert 0\right \rangle }_{\text{M}}$.
The braiding
operator is obtained as $\mathcal{R}_{\text{M}}^{\beta}=e^{-i\Delta \Phi \tau_{z}}$ which
is the NH generalization of the Ivanov's braiding operator. Here, $\tau_{z}$
denotes a Pauli matrix on the Majorana qubit $(\left \vert 0\right \rangle
_{\text{M}}^{\beta},\left \vert 1\right \rangle _{\text{M}}^{\beta})$. $\Delta \Phi=\frac{1}{2}(\Delta \phi_{1}^{\beta}-\Delta \phi_{0}^{\beta})=\frac{\pi}{8}+\frac{\pi
}{4(e^{-4\beta}+1)}$ denotes a Berry phase during braiding
processes that can \emph{continuously} tuned from $\pi/8$ to
$3\pi/8$. Thus, $\Delta \Phi$ can be an arbitrary value in the region
of $(\pi/8,3\pi/8)$, including rational number or
irrational number. As a result, we call it \emph{irrational non-Abelian
statistics}. By contrast, the Berry phase from braiding processes for usual
non-Ableian statistics is fixed to $\Delta \Phi=\pi/4.$ Besides, when we fix $\beta$,
for two non-Hermitian Majorana zero modes far away, the braiding rule and the
corresponding Berry phase $\Delta \Phi$ will never change, no matter how you
change the braiding path! In this sense, this is a remarkable example of
``\emph{irrational topological phenomenon}" and can be considered as
non-Abelian generalization of irrational Abelian statistics for \textrm{U(1)}
Abelian anyons according to Wilczek flux-binding picture!

\textbf{Example for numerical simulations on verifying the irrational
non-Abelian statistics.} A 1D NH Kitaev model \cite{kitaev2001} with
imbalanced $p$-wave SC paring is taken as an example to illustrate the
anomalous non-Abelian statistics of NH MZMs, and the numerical simulations are
performed during the braiding processes. The Hamiltonian is written as
\begin{align}
\mathrm{\hat{H}}_{\mathrm{NHK}}(\beta)  &  =-\sum_{j=1}^{N}[t(c_{j}^{\dagger
}c_{j+1}+c_{j+1}^{\dagger}c_{j})+\Delta^{+}c_{j}^{\dagger}c_{j+1}^{\dagger
}\nonumber \\
&  +\Delta^{-}c_{j+1}c_{j}+\mu(1-2n_{j})],
\end{align}
where $c_{j}$ $(c_{j}^{\dagger})$ annihilates (creates) a fermion on site $j$.
$t,$ $\Delta^{\pm},$ $\mu$ and $N$ denote the hopping amplitude, the strength
of $p$-wave pairing, the chemical potential and the lattice number,
respectively. We set $\Delta^{\pm}=\Delta_{0}e^{\pm2\beta}$, where $\beta
\in \mathbb{R}$ represents the NH strength and $\Delta_{0}>0$. When $\beta
\neq0,$ we have $\mathrm{\hat{H}}_{\mathrm{NHK}}\neq \mathrm{\hat{H}%
}_{\mathrm{NHK}}^{\dagger}$, which can be achieved by the NH similarity
transformation from it's Hermitian counterpart. In this letter, we focus on
the case of $t=\Delta_{0}$.

The 1D NH SC may have nontrivial topological properties (see Appendix-B for
details). For the translation variables ansatz, we transform the fermion
Hamiltonian into momentum space, $\mathrm{\hat{H}}_{\mathrm{NHK}}(k)=\sum
_{k}\psi_{k}^{\dag}h(k,\beta)\psi_{k}$ with
\begin{equation}
h(k,\beta)=(t\cos k+\mu)\cdot \sigma^{z}+\Delta_{0}\sin k\cdot \sigma^{y,\beta}%
\end{equation}
by introducing $\psi_{k}=(c_{k},c_{-k}^{\dag})^{T}$, where $\sigma
_{j}^{y,\beta}=\mathcal{S}\sigma_{j}^{y}\mathcal{S}^{-1}=\cosh(\beta
)\sigma_{j}^{y}-i\sinh(\beta)\sigma_{j}^{x}$ is a $2\times2$ matrix. With the
help of the biorthogonal set, we define right/left eigenstates for the NH
systems as $\mathrm{\hat{H}}_{\mathrm{NHK}}|{\Psi}_{n}^{\mathrm{R}}%
\rangle=E_{n}|{\Psi}_{n}^{\mathrm{R}}\rangle,$ and $\mathrm{\hat{H}%
}_{\mathrm{NHK}}^{\dagger}|{\Psi}_{n}^{\mathrm{L}}\rangle=E_{n}^{\ast}|{\Psi
}_{n}^{\mathrm{L}}\rangle,$ where $E_{n}$, $E_{n}^{\ast}$ are the
corresponding eigenvalues (with $n=0,1$ representing the two lowest energy
states). To describe the topological structure of $\mathrm{\hat{H}%
}_{\mathrm{NHK}}$, we define \emph{biorthogonal }$\mathrm{Z}_{2}$\emph{
topological invariant},
\begin{equation}
\mathcal{\omega}=\mathrm{sgn}(\eta_{_{k=0}}\cdot \eta_{_{k=\pi}})
\end{equation}
where $\eta_{_{k=0/\pi}}={}\left \langle {\Psi}_{0}^{\mathrm{L}}\right \vert
c_{k=0/\pi}^{\dagger}c_{k=0/\pi}\left \vert {\Psi}_{0}^{\mathrm{R}%
}\right \rangle {}$and $\eta_{_{k=0/\pi}}(\beta)=\eta_{_{k=0/\pi}}(\beta=0)$.
Therefore, we have $\eta_{_{k=0}}=\mathrm{sgn}(t+\mu),$ $\eta_{_{k=\pi}%
}=\mathrm{sgn}(-t+\mu).$ For the case of $\mathcal{\omega}=1$ ($\left \vert
t\right \vert <\left \vert \mu \right \vert $),\ the SC is trivial; But for
$\mathcal{\omega}=-1$ ($\left \vert t\right \vert >\left \vert \mu \right \vert
$),\ the SC becomes topological one.

In the topological phase with $\mathcal{\omega}=-1$, there exist two edge
states with (nearly) zero energy, i.e., the NH MZMs at the left end
$\gamma_{L}^{\beta}$ and at the right end $\gamma_{R}^{\beta}$. After defining
the fermionic operators $\tilde{C}_{\mathrm{M}}=(\gamma_{L}^{\beta}%
+i\gamma_{R}^{\beta})/2,$ $\tilde{C}_{\mathrm{M}}^{\dagger}=(\gamma_{L}%
^{\beta}-i\gamma_{R}^{\beta})/2$, we obtain the two ground states with the
open boundary condition as $\left \vert {\Psi}_{0}(\beta)\right \rangle
=\tilde{C}_{\mathrm{M}}\left \vert F\right \rangle $, $\left \vert {\Psi}%
_{1}(\beta)\right \rangle =\tilde{C}_{\mathrm{M}}^{\dagger}\left \vert {\Psi
}_{0}(\beta)\right \rangle ,$ where $\left \vert F\right \rangle $ is the NH
ground state with occupied single particle states for $E<0$ and empty single
\begin{figure}[t]
\centering
\includegraphics[scale=0.27]{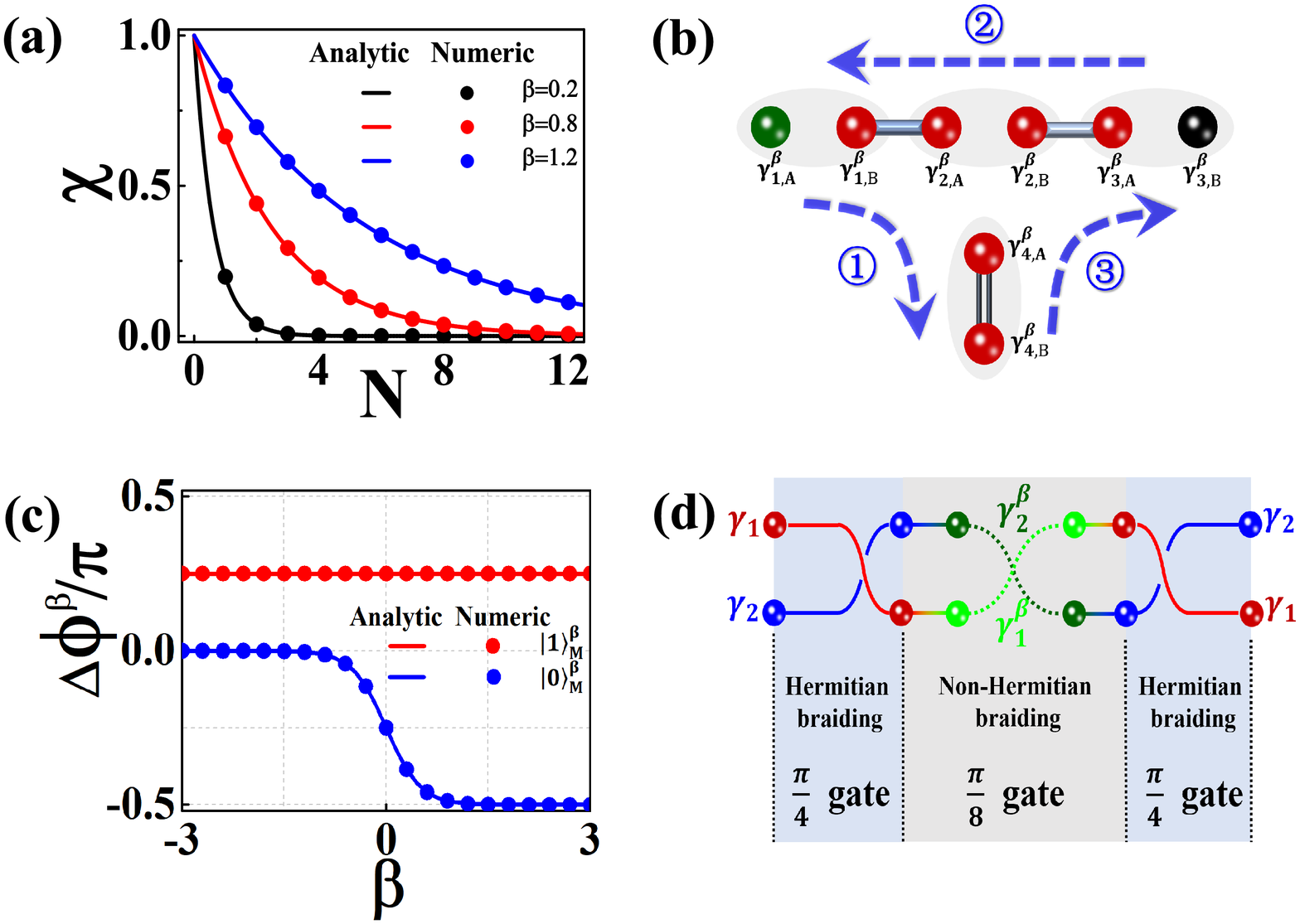}
\caption{(a) The numerical results (dots) and the analytical results (lines)
for the similarity between two degenerate ground states in NH Kitaev model
with $t=\Delta_{0},$ $\mu=0$, and $\beta=0.2,$ $0.8$ and $1.2$. These results
indicate the orthogonality of two degenerate ground states in thermodynamic
limit ($N\mapsto \infty$); (b) Schematic diagram for the T-type braiding
process to exchange the two NH MZMs. We take a system with 8 Majorana fermions
as an example; (c) The Berry phase for the quantum states $\left \vert
0\right \rangle _{\text{M}}^{\beta}$ and $\left \vert 1\right \rangle _{\text{M}%
}^{\beta}$ during the braiding processes. (d) An illustration of NH assisted
TQC. In step-2, a $\pi/8$ gate is realized by tuning the NH strength $\beta$,
dotted lines indicate that multiple braiding operations can be performed.}%
\end{figure}particle states for $E\geq0$. We find that in the thermodynamic
limit ($N\mapsto \infty$) the energy splitting of the two MZMs is zero and
$\left \vert {\Psi}_{0}(\beta)\right \rangle $ and $\left \vert {\Psi}_{1}%
(\beta)\right \rangle $ are orthogonal by calculating the similarity between
them, which is defined as $\chi(\beta)=\left \vert \langle{\Psi}_{0}%
(\beta)|{\Psi}_{1}(\beta)\rangle \right \vert $. Here, $|{\Psi}_{0/1}%
(\beta)\rangle$ satisfies the self-normalization condition $|\langle{\Psi
}_{0/1}(\beta)|{\Psi}_{0/1}(\beta)\rangle|=1$. For example, when $t=\Delta
_{0}$ and $\mu=0$ we have $\chi(\beta)=(\tanh \beta)^{N}$. It is obvious that
$\chi(\beta)\mapsto0$ with $N\mapsto \infty$, so the two degenerate ground
states are orthogonal. The proof of the orthogonal properties is shown in
Fig.2(a) where the numeric results (the dots) are consisted with the analytic
results (the lines).

According to the definition, the fermion parity of $\left \vert {\Psi}%
_{0}(\beta)\right \rangle $ is even and the fermion parity of $\left \vert
{\Psi}_{1}(\beta)\right \rangle $ is odd. Therefore, due to the orthogonality
and the parities of the two-fold degenerate ground states, we can use them to
construct the two basis states of the NH Majorana qubit (see the Appendix-D1).
We introduce the NH Majorana qubit in this system as: $\left \vert
0\right \rangle _{\text{M}}^{\beta}\equiv \left \vert {\Psi}_{0}(\beta
)\right \rangle =\tilde{C}_{\mathrm{M}}\left \vert {F}\right \rangle ,$
$\left \vert 1\right \rangle _{\text{M}}^{\beta}\equiv \left \vert {\Psi}%
_{1}(\beta)\right \rangle =\tilde{C}_{\mathrm{M}}^{\dagger}\left \vert
0\right \rangle _{\text{M}}^{\beta}.$

The non-Abelian statistics of two NH MZMs can be verified in the T-junction
Majorana chain systems \cite{Alicea2011,T-Type2}, which contain $4$ lattice
sites, as shown in Fig.2(b). Here, the braiding processes of the two NH MZMs
are denoted by blue dotted arrows. We perform the numerical simulations to
verify the non-Hermitian Ivanov's braiding operator $\mathcal{R}_{\text{M}%
}^{\beta}$ for two NH MZMs by mapping the original fermionic model
$\mathrm{\hat{H}}_{\mathrm{NHK}}(\beta)$ onto a NH transverse Ising model via
the NH Jordan-Wigner transformation \cite{Guo2016}(see Appendix-C). As a
result, the Hamiltonian becomes
\begin{equation}
\mathrm{\hat{H}}_{\mathrm{NHK}}(\beta)=-\frac{1}{4}\sum_{j}(J{\sigma}%
_{j}^{x,\beta}\sigma_{j+1}^{x,\beta}-4h\sigma_{j}^{z}) \label{Ising}%
\end{equation}
where $\sigma_{j}^{x,\beta}=\mathcal{S}(\beta)\sigma_{j}^{x}\mathcal{S}%
^{-1}(\beta)=\cosh(\beta)\sigma_{j}^{x}+i\sinh(\beta)\sigma_{j}^{y}$ and
$S(\beta)=\mathrm{exp}(\frac{\beta}{2}\sum_{i}\sigma_{i}^{z}),J=t=\Delta
_{0},h=\mu$. The non-Hermitian model in Eq.(\ref{Ising}) can be simulated
using three-level atoms in a variety of setups (see Appendix-E), including
trapped ions, cavity QED, and atoms in optical lattices. The dynamics by
$H(\beta)$ can be decomposed as
\begin{equation}
e^{-iH(\beta)t}=e^{-i\mu \sigma^{z}t}S(\beta)(\prod_{i}e^{i\frac{J}{4}{\sigma
}_{j}^{x}\sigma_{j+1}^{x}t})S^{-1}(\beta),
\end{equation}
where the nonunitary dynamics $S(\beta)$ and $S^{-1}(\beta)$ are from
measuring whether a spontaneous decay has occurred
\cite{betaEx1,betaEx2,betaEx3}. This process can be measured with a high
degree of accuracy \cite{Exp2006,Exp2008,Exp2013}.

Meanwhile, the braiding process for the Majorana qubit ($\left \vert
0\right \rangle _{\text{M}}^{\beta},\left \vert 1\right \rangle _{\text{M}%
}^{\beta}$) is mapped onto that for the two degenerate ground states in the
spin representation. The irrational Ivanov's braiding operator $\mathcal{R}%
_{\text{M}}^{\beta}$ for two MZMs is mapped onto the corresponding operator
$R^{z}(\varphi)$, which rotate the spin $\varphi$ angle around the $z$-axis in
spin representation, i.e., $\mathcal{R}_{\text{M}}^{\beta}\leftrightarrow
R^{z}(\varphi)$. This is the same as the case in the Hermitian system. The
Berry phases for the quantum states $\left \vert 0\right \rangle _{\text{M}%
}^{\beta}$ and $\left \vert 1\right \rangle _{\text{M}}^{\beta}$ in the Majorana
braiding process are calculated by the Wilson loop method like Eq.\ref{Wloop}.
The numerical results (dots) are exactly consistent with the theoretical
prediction (lines) as shown in Fig.2(c). In supplementary materials, we show
detailed discussion on numerical simulations and theoretical derivation of the
Ivanov's braiding processes(see Appendix-D2,D3).

\textbf{Non-Hermitian assisted topological quantum computation via
Non-Hermitian MZMs.} Due to non-locality and orthogonality, the NH MZMs may be
utilized as a decoherence-free qubit, which play an important role in the
realization of fault-tolerant universal TQC. We propose an alternative
approach to universal TQC via NH MZMs -- \emph{Non-Hermitian assisted TQC}.

If one can realize $\mathrm{\hat{H}}_{\mathrm{NHK}}^{\beta}$ with the freely
adjustable NH strength $\beta$, we can adiabatically tune $\beta$ to construct
a universal TQC. For the Hadamard gate, phase gate, and controlled NOT gate,
we set the NH strength $\beta$ to zero. For the $\pi/8$ gate, we set the NH
strength $\beta$\ to be certain value and braid NH MZM for $\mathcal{N}$
times. For example, $\mathcal{N}=4$ for $\beta=-(\mathrm{ln}0.6)/4\approx
0.128$. For this case, during the braiding processes, the $\pi/8$ gate can be
reached $\mathbf{T}=[\mathcal{R}_{\text{M}}^{\beta}]^{\mathcal{N}}$. In the
end, to perform measurement, the NH strength $\beta$ returns to zero again.
What should be mentioned is that $\mathbf{T}$-gate from braiding process is
based on ``\emph{irrational topological phenomenon}", which has huge
advantages over other non-topological methods, such as the method of ``magic
state distillation" \cite{QbitKitaev2005,QbitDsarma2005}.

In Fig.2(d), an illustration of two phase gates $S$ and a $\pi/8$ gate for NH
assisted TQC is shown. Here, we take a braiding process with three steps as an
example: a phase gate $S$ by exchanging two Hermitian MZMs with $\beta=0$, a
$\pi/8$ gate by exchanging two NH MZMs with $\beta \neq0,$ and a phase gate $S$
by exchanging two MZMs with $\beta=0.$

\textbf{Conclusion and discussion:} In this letter, we developed a theory for
NH generalization of MZMs, i.e., $\gamma^{\beta}=\mathcal{S}\gamma
\mathcal{S}^{-1}$ where $\mathcal{S}$ is the NH PH similarity transformation
and $\beta$ is the NH strength. The key point of NH generalization of MZMs is
\emph{irrational non-Abelian statistics}, an example for ``\emph{irrational
topological phenomenon}". Due to the particle-hole-symmetry breaking, the
Berry phase from braiding processes become an arbitrary number in a region,
i.e., $\Delta \Phi \in(\pi/8,3\pi/8)$. The irrational non-Abelian statistics for
the NH MZMs indicates that in NH topological systems the theory for usual
unitary modular tensor category would be generalized to a theory for certain
non-unitary modular tensor category, and the theory for usual topological
field theories would be generalized to a theory for certain non-Hermitian
topological field theories. In the future, we will study these issues. In
addition, we plan to apply the theory to other TSCs, such as the 2D NH
$p_{x}+ip_{y}$ TSC and higher-order NH TSCs, and then study the possible
physical realization of the NH MZMs in these NH topological systems.

\acknowledgments This work is supported by NSFC Grant No. 1217040237,
11974053, 61835013, National Key R$\&$D Program of China under grants No.
2016YFA0301500, Strategic Priority Research Program of the Chinese Academy of
Sciences under grants Nos. XDB01020300, XDB21030300.


\begin{thebibliography}{99}                                                                                               %


\bibitem {Ivanov2001}D. A. Ivanov, Phys. Rev. Lett. \textbf{86}, 268 (2001).


\bibitem {read2000}N. Read, and D. Green, Phys. Rev. B \textbf{61}, 10267
(2000).


\bibitem {kitaev2001}A. Y. Kitaev, Phys. Usp. \textbf{44}, 131 (2001).


\bibitem {kitaev1997}A. Y. Kitaev, Russ. Math.Surv. 52, 1191 (1997).


\bibitem {Sarma2006}S. D. Sarma, C. Nayak, and S. Tewari, Phys. Rev. B
\textbf{73}, 220502 (2006).


\bibitem {Tewari2007}B. Lian, X. Q. Sun, A. Vaezi, X. L. Qi, and S. C. Zhang,
Proc. Natl. Acad. Sci. U.S.A. \textbf{115}, 10938 (2018).


\bibitem {Nayak2008}C. Nayak, S. H. Simon, A. Stern, M. Freedman, and S. D.
Sarma, Rev. Mod. Phys. \textbf{80}, 1083 (2008).


\bibitem {Fu2008}L. Fu, and C. Kane, Phys. Rev. Lett. \textbf{100}, 096407
(2008).


\bibitem {Stern2010}A. Stern, Nature, \textbf{464}, 187 (2010).


\bibitem {Sau2010}J. D. L. Sau, R. M. Tewar, and S. D. Sarma, Phys. Rev. Lett.
\textbf{104}, 040502 (2010).


\bibitem {Alicea2011}J. Alicea, Y. Oreg, G. Refael, F. von Oppen, and M. P. A.
Fisher, Nat. Phys. \textbf{7}, 412 (2011).


\bibitem {Mourik2012}V. Mourik, K. Zuo, S. M. Frolov, S. R. Plissard, E. P. A.
M. Bakkers, L. P. Kouwenhoven, Science \textbf{336}, 1003 (2012).


\bibitem {Deng2012}M. T. Deng, C. L. Yu, G. Y. Huang, M. Larsson, P. Caroff,
H. Q. Xu, Nano. Lett. \textbf{12}, 6414 (2012).


\bibitem {Rokhinson2012}L. P. Rokhinson, X. Liu, and J. K. Furdyna, Nat. Phys.
\textbf{8}, 795 (2012).


\bibitem {Alicea2012}J. Alicea, Rep. Prog. Phys. \textbf{75}, 076501 (2012)


\bibitem {Mebrahtu2013}H. T. Mebrahtu, I. V. Borzenets, H. Zheng, Y. V. Bomze,
A. I. Smirnov, S. Florens, H. U. Baranger \and G. Finkelstein, Nat. Phys.
\textbf{9}, 732 (2013).


\bibitem {Nadj-Perge2014}S. Nadj-Perge, et al. Science \textbf{346}, 602
(2014).


\bibitem {Lee2014}E. J. H. Lee, X. Jiang, M. Houzet, R. Aguado, C. M. Lieber,
\and S. D. Franceschi, Nat. Nano \textbf{9}, 79 (2014).






\bibitem {Rudner2009}M. S. Rudner and L. S. Levitov, Phys. Rev. Lett.
\textbf{102}, 065703 (2009).


\bibitem {Esaki2011}K. Esaki, M. Sato, K. Hasebe, and M. Kohmoto, Phys. Rev. B
\textbf{84}, 205128 (2011).


\bibitem {Hu2011}Y. C. Hu and T. L. Hughes, Phys. Rev. B \textbf{84}, 153101
(2011).


\bibitem {Liang2013}S. D. Liang and G. Y. Huang, Phys. Rev. A \textbf{87},
012118 (2013).


\bibitem {Zhu2014}B. Zhu, R. L\"{u}, and S. Chen, Phys. Rev. A \textbf{89},
062102 (2014).




\bibitem {Zeuner2015}J. M. Zeuner, M. C. Rechtsman, Y. Plotnik, Y. Lumer, S.
Nolte, M. S. Rudner, M. Segev, and A. Szameit, Phys. Rev. Lett. \textbf{115},
040402 (2015).


\bibitem {Wang2015}X. Wang, T. Liu, Y. Xiong, et al. Phys. Rev. A \textbf{92},
012116 (2015).


\bibitem {Yuce2016}C. Yuce, Phys. Rev. A \textbf{93}, 062130 (2016).


\bibitem {Zeng2016}Q. B. Zeng, B. Zhu, S. Chen, L. You, R. L\"{u}, Phys. Rev.
A \textbf{94}, 022119 (2016).


\bibitem {Menke2017}H. Menke, M. M. Hirschmann, Phys. Rev. B \textbf{95},
174506 (2017).


\bibitem {Kawabata2018}K. Kawabata, Y. Ashida, H. Katsura and M. Ueda, Phys.
Rev. B \textbf{98}, 085116 (2018).


\bibitem {Lieu2019}S. Lieu, Phys. Rev. B \textbf{100}, 085110 (2019).


\bibitem {Li2018}C. Li, X. Z. Zhang, G. Zhang, et al. Phys. Rev. B
\textbf{97}, 115436 (2018).


\bibitem {Lee2016}T. E. Lee, Phys. Rev. Lett. \textbf{116}, 133903 (2016).


\bibitem {San2016}P. S. Jose, J. Cayao, E. Prada, and R. Aguado, Sci. Rep.
\textbf{6}, 21427 (2016).




\bibitem {Weimann2017}S. Weimann, M. Kremer, Y. Plotnik, Y. Lumer, S. Nolte,
K. G. Makris, M. Segev, M. C. Rechtsman, and A. Szameit, Nat. Mater.
\textbf{16}, 433 (2017).


\bibitem {Xiao2017}L. Xiao, X. Zhan, Z. H. Bian, K. K. Wang, X. Zhang, X. P.
Wang, J. Li, K. Mochizuki, D. Kim, N. Kawakami, W. Yi, H. Obuse, B. C.
Sanders, and P. Xue, Nat. Phys. \textbf{13}, 1117 (2017).


\bibitem {Leykam2017}D. Leykam, K. Y. Bliokh, C. Huang, Y. D. Chong, and F.
Nori, Phys. Rev. Lett. \textbf{118}, 040401 (2017).


\bibitem {Shen2018}H. Shen, B. Zhen, and L. Fu, Phys. Rev. Lett. \textbf{120},
146402 (2018).


\bibitem {Lieu2018}S. Lieu, Phys. Rev. B \textbf{97}, 045106 (2018).


\bibitem {Xiong2018}Y. Xiong, J. Phys. Commun. \textbf{2}, 035043 (2018).


\bibitem {Gong2018}Z. Gong, Y. Ashida, K. Kawabata, K. Takasan, S.
Higashikawa, and M. Ueda, Phys. Rev. X \textbf{8}, 031079 (2018).


\bibitem {Yao2018}S. Yao, and Z. Wang, Phys. Rev. Lett. \textbf{121}, 086803
(2018).


\bibitem {YaoWang2018}S. Yao, F. Song, and Z. Wang, Phys. Rev. Lett.
\textbf{121}, 136802 (2018).


\bibitem {Kunst2018}F. K. Kunst, E. Edvardsson, J. C. Budich, and E. J.
Bergholtz, Phys. Rev. Lett. \textbf{121}, 026808 (2018).


\bibitem {Yin2018}C. Yin, H. Jiang, L. Li, R. L\"{u}, and S. Chen, Phys. Rev.
A \textbf{97}, 052115 (2018).


\bibitem {Borgnia2020}D. S. Borgnia, A. J. Kruchkov, and R. J. Slager Phys.
Rev. Lett. \textbf{124}, 056802 (2018)

\bibitem {KawabataUeda2018}K. Kawabata, K. Shiozaki, and M. Ueda, Phys. Rev. B
\textbf{98}, 165148 (2018).


\bibitem {Alvarez2018}V. M. M. Alvarez, J. E. B. Vargas, M. Berdakin, and L.
E. F. F. Torres, Eur. Phys. J. Spec. Top. \textbf{227}, 1295 (2018).


\bibitem {Jiang2018}H. Jiang, C. Yang, and S. Chen, Phys. Rev. A \textbf{98},
052116 (2018).




\bibitem {Bandres2018}M. A. Bandres, S. Wittek, G. Harari, M. Parto, J. Ren,
M. Segev, D. N. Christodoulides, and M. Khajavikhan, Science \textbf{359},
4005 (2018).


\bibitem {Zhou20182}H. Zhou, C. Peng, Y. Yoon, C. W. Hsu, K. A. Nelson, L. Fu,
J. D. Joannopoulos, M. Soljacic, and B. Zhen, Science \textbf{359}, 1009
(2018).


\bibitem {Cerjan2019}A. Cerjan, S. Huang, M. Wang, K. P. Chen, Y. Chong, and
M. C. Rechtsman, Nat. Photon. \textbf{13}, 623 (2019).


\bibitem {Wang2019}K. Wang, X. Qiu, L. Xiao, X. Zhan, Z. Bian, B. C. Sanders,
W. Yi, and P. Xue, Nat. Commun. \textbf{10}, 2293 (2019).


\bibitem {Xiaoxue2019}L. Xiao, T. Deng, K. Wang, G. Zhu, Z. Wang, W. Yi, P.
Xue, Nat. Phys. \textbf{16}, 761 (2020).


\bibitem {Helbig2019}T. Helbig, T. Hofmann, S. Imhof, M. Abdelghany, T.
Kiessling, L. W. Molenkamp, C. H. Lee, A. Szameit, M. Greiter, and R. Thomale,
Nat. Phys. \textbf{16}, 747 (2020)


\bibitem {Ghatak2019}A. Ghatak and T. Das, J. Phys.: Condens. Matter
\textbf{31}, 263001 (2019).


\bibitem {Avila2019}J. Avila, F. Pe\~{n}randa, E. Prada, P. San-Jose, and R.
Aguado, Commun. Phys. \textbf{2}, 1 (2019).


\bibitem {Jin2019}L. Jin and Z. Song, Phys. Rev. B \textbf{99}, 081103
(2019);S. Lin, L. Jin, and Z. Song, Phys. Rev. B \textbf{99}, 165148 (2019);
K. L. Zhang, H. C. Wu, L. Jin, and Z. Song, Phys. Rev. B \textbf{100}, 045141
(2019).


\bibitem {Lee2019}C. H. Lee and R. Thomale, Phys. Rev. B \textbf{99},
201103(R) (2019).


\bibitem {Liu2019}T. Liu, Y. R. Zhang, Q. Ai, Z. Gong, K. Kawabata, M. Ueda,
and F. Nori, Phys. Rev. Lett. \textbf{122}, 076801 (2019).


\bibitem {38-1}K. Kawabata, K. Shiozaki, M. Ueda, and M. Sato, Phys. Rev. X
\textbf{9}, 041015 (2019).


\bibitem {38}H. Zhou and J. Y. Lee, Phys. Rev. B \textbf{99}, 235112 (2019).


\bibitem {chen-class2019}C. H. Liu, H. Jiang, S. Chen, Phys. Rev. B
\textbf{99}, 125103 (2019).


\bibitem {Edvardsson2019}E. Edvardsson, F. K. Kunst, and E. J. Bergholtz,
Phys. Rev. B \textbf{99}, 081302(R) (2019).


\bibitem {Herviou2019}L. Herviou, J. H. Bardarson, and N. Regnault, Phys. Rev.
A \textbf{99,} 052118 (2019).


\bibitem {Yokomizo2019}K. Yokomizo and S. Murakami, Phys. Rev. Lett.
\textbf{123}, 066404 (2019).


\bibitem {Kunst2019}F. K. Kunst and V. Dwivedi, Phys. Rev. B \textbf{99},
245116 (2019).


\bibitem {zhouBin2019}R. Chen, C. Z. Chen, B. Zhou, and D. H. Xu, Phys. Rev. B \textbf{99}, 155431 (2019).


\bibitem {Deng2019}T. S. Deng and W. Yi, Phys. Rev. B \textbf{100}, 035102
(2019).


\bibitem {SongWang2019}F. Song, S. Yao, and Z. Wang, Phys. Rev. Lett. \textbf{123}, 170401 (2019).


\bibitem {xi2019}X. W. Luo and C. W. Zhang, Phys. Rev. Lett. \textbf{123}
073601 (2019).


\bibitem {Longhi2019}S. Longhi, Phys. Rev. Research \textbf{1}, 023013
(2019).


\bibitem {chen-edge2019}H. Jiang, R. L\"{u}, S Chen, Eur. Phys. J. B \textbf{93}, 125 (2020).


\bibitem {Ashida2020}Y. Ashida, Z. Gong, and M. Ueda, Adv. Phys. \textbf{69}, 3 (2020).


\bibitem {T-Type2}X. M. Zhao, J. Yu, J. He, Q. B. Cheng, Y. Liang, S. P. Kou.
Mod. Phys. Lett. B. \textbf{31},1750123 (2017)


\bibitem {PhaseGate}Based on the Solovay-Kitaev algorithm reval in
Ref.\cite{kitaev1997}, to realize universal TQC we just need to construct a
arbitrary phase gate $\mathbf{P}=\mathrm{diag}\{e^{i\Delta \phi},e^{-i\Delta
\phi}\}$ with phase changing $\Delta \phi \neq0,$ $\pm \pi/4,$ $\pm \pi/2,$ $\pi$
but not must be fixed to $\pi/8$. As a result, the gate with an arbitrary
(irrational) phase changing can be realized by finite $\beta$ but not must be
fixed to $\beta \rightarrow-\infty$.

\bibitem {QbitKitaev2005}S. Bravyi and A. Kitaev, Phys. Rev. A \textbf{71},
022316 (2005)

\bibitem {QbitDsarma2005}S. D. Sarma, M. Freedman and C. Nayak, npj Quantum
Inf \textbf{1}, 15001 (2015).

\bibitem {betaEx1}T. E. Lee and C. K. Chan, Phys. Rev. X \textbf{4}, 041001 (2014).

\bibitem {betaEx2}T. E. Lee, F. Reiter, and N. Moiseyev, Phys. Rev. Lett.
\textbf{113}, 250401 (2014)

\bibitem {betaEx3}H. Weimer, M. Muller, I. Lesanovsky, P. Zoller, and H. P.
Buchler, Nat. Phys. \textbf{6}, 382 (2010).

\bibitem {Exp2006}N. Katz, M. Ansmann, R. C. Bialczak, E. Lucero, R.
McDermott, M. Neeley, M. Steffen, E. M. Weig, A. N. Cleland, J. M. Martinis,
and A. N. Korotkov, Science \textbf{312}, 1498 (2006).

\bibitem {Exp2008}A. H. Myerson, D. J. Szwer, S. C. Webster, D. T. C. Allcock,
M. J. Curtis, G. Imreh, J. A. Sherman, D. N. Stacey, A. M. Steane, and D. M.
Lucas, Phys. Rev. Lett. \textbf{100}, 200502 (2008).

\bibitem {Exp2013}J. A. Sherman, M. J. Curtis, D. J. Szwer, D. T. C. Allcock,
G. Imreh, D. M. Lucas, and A. M. Steane, Phys. Rev. Lett. \textbf{111}, 180501 (2013).

\bibitem {Guo2016}J. S. Xu, K. Sun, Y. J. Han, C. F. Li, J. K. Pachos, and G.
C. Guo, Nat. Commun. \textbf{7},13194 (2016)

\end{thebibliography}
\end{document}